\documentclass[fleqn,11pt]{article}

\usepackage{amsmath,amsfonts,amssymb}
\usepackage{color}

\usepackage{geometry}
\newtheorem{theorem}{Theorem}
\newtheorem{lemma}[theorem]{Lemma}

\newtheorem{definition}{Definition}

\newcommand{\R}{\mathbb{R}}

\newcommand{\N}{\mathbb{N}}
\newcommand{\T}{\mathbb{T}}

\newcommand{\cqfd}
{%
\mbox{}%
\nolinebreak%
\hfill%
\rule{2mm}{2mm}%
\newline
\newline
}

\title{On the Cauchy problem with large data for a space-dependent Boltzmann-Nordheim boson equation II.}
\author{Leif ARKERYD and Anne NOURI\\
\\Mathematical Sciences, 41296 G\"oteborg, Sweden,\\
arkeryd@chalmers.se\\
Aix-Marseille University, CNRS, Centrale Marseille, I2M UMR 7373, 13453 Marseille, France,\\
anne.nouri@univ-amu.fr}

\date{}

\begin{document}

\maketitle

{\noindent \bf Abstract.}\hspace{0.1in}
This paper studies a space-inhomogenous Boltzmann Nordheim equation with pseudo-Maxwellian forces. Strong  solutions are obtained for the Cauchy problem with large initial data in an $L^1\cap L^\infty$ setting. The main results are existence, uniqueness, stability and qualitative $L^\infty $ features of solutions conserving mass, momentum and energy.

\footnotetext[2]{Key words; bosonic Boltzmann-Nordheim equation, low temperature kinetic theory, quantum Boltzmann equation.}
%
%
%
%
\section{Introduction and main result.}
In a previous paper \cite{AN1}, we have studied the Cauchy problem for the Boltzmann-Nordheim {\cite{N}} boson equation in a slab with two-dimensional velocity space,
\begin{equation}\label{f-bn2D}
\partial _tf(t,x,v)+v_1\partial _xf(t,x,v)= Q_{0}(f)(t,x,v),
\quad f(0,x ,v )= f_0(x,v) \nonumber
,\hspace*{0.02in} {(t,x)\in \R _+\times [ 0,1] ,\hspace*{0.02in} v=(v_1,v_2)\in \R ^2.}
\end{equation}
Its solution was obtained as the limits when $\alpha \rightarrow 0$ of the kinetic equation for anyons,
\begin{equation}\label{f-alpha}
\partial _tf(t,x,v)+v_1\partial _xf(t,x,v)= Q_{\alpha }(f)(t,x,v),
\quad f(0,x ,v )= f_0(x,v)
,\hspace*{0.02in} {(t,x)\in \R _+\times [ 0,1] ,\hspace*{0.02in} v=(v_1,v_2)\in \R ^2.} \nonumber
\end{equation}
The collision operator $Q_{\alpha }$ in \cite{AN2} depends on a parameter $\alpha \in [ 0,1] $ and is given by
\begin{eqnarray*}
Q_\alpha (f)(v)= \int_{I\! \!R^2 \times S^{1}}B(|v-v_*|,n)
 [f^\prime f^\prime _*F_\alpha(f)F_\alpha(f_*)-ff_*F_\alpha(f^\prime )F_\alpha(f^\prime _*)] dv_*dn,\hspace{.1cm}
\end{eqnarray*}
with the kernel B of Maxwellian type, $f^\prime $, $f^\prime _*$, $f$, $f_*$ the values of $f$ at $v^\prime $, $v^\prime _*$, $v$ and $v_*$ respectively, where
\begin{align*}
v^\prime = v-(v-v_*, n)n ,\quad v^\prime _*= v_*+(v-v_*, n)n \, ,
\end{align*}
and the filling factor $F_\alpha$
\begin{eqnarray*}
F_\alpha(f)= (1-\alpha f)^{\alpha}(1+(1-\alpha)f)^{1-\alpha}\, .
\end{eqnarray*}
In this paper, we solve in a direct way the Cauchy problem for the three-dimensional Boltzmann-Nordheim equation in a torus,
\begin{equation}\label{f-bn3D}
\partial _tf(t,x,v)+v\cdot \nabla _xf(t,x,v)= R_{0}(f)(t,x,v),
\quad f(0,x ,v )= f_0(x,v)
,\hspace*{0.02in} {(t,x,v)\in \R _+\times \T ^3\times \R ^3},
\end{equation}
where
\begin{equation}\label{df-R0}
R_0(f)(v)= \int_{I\! \!R^3 \times S^{2}}B(|v-v_*|,n)
 [f^\prime f^\prime _*(1+f)(1+f_*)-ff_*(1+f^\prime )(1+f^\prime _*)] dv_*dn.\nonumber
\end{equation}
\hspace*{1.in}\\
The Boltzmann-Nordheim equation (\ref{f-bn3D}) was initiated by Nordheim \cite{N}, Uehling and Uhlenbeck \cite{U} using quantum statistical considerations. All quantum features appear at the level of the collision operator $R_0$. For a gas of bosons, the quantum effects are taken into account by
the probability of collision between two particles depending on the number of particles occupying the state after collision. Solutions to (\ref{f-bn3D}) satisfy an entropy principle and equilibrium states are the following entropy minimizers
\begin{equation}\label{equilibrium}
\frac{1}{e^{\frac{\lvert v-u\rvert ^2-\mu }{2T}}-1}+m_0\delta _{v-u},
\end{equation}
where $u\in R^3$, $\mu \leq 0$ is the chemical potential, $m_0\geq 0$ and $\mu \hspace*{0.01in}m_0= 0$. It is expected that a global in time solution to (\ref{f-bn3D}) should converge to the equilibrium state (\ref{equilibrium}) with the same mass, momentum and kinetic energy as its initial datum. This gives rise to a critical kinetic temperature $T_c$
such that the initial distribution $m_0$ is different from zero if and only if $T<T_c$. {It is a reason why one should only expect local in time existence results in $L^\infty $ for (\ref{f-bn3D}), if no restriction on the temperature of the initial datum is made.}\\
For the bosonic BN equation general existence results were first obtained by X. Lu in \cite{Lu1} in the space-homogeneous isotropic large data case. It was followed by a number of interesting studies in the same isotropic setting, by X. Lu \cite{Lu2,Lu3,Lu4}, and by M. Escobedo and J.L. Vel\'azquez \cite{EV2,EV}. Results with the isotropy assumption removed, were recently obtained by M. Briant and A. Einav \cite{BE}. Finally a space-dependent case close to equilibrium has been studied by G. Royat in \cite{R}.\\
The papers \cite{Lu1,Lu2,Lu3,Lu4} by Lu, study the isotropic, space-homogeneous BN equation both for Cauchy data leading to mass and energy conservation, and for data leading to mass loss when time tends to infinity. Escobedo and Vel\'asquez in \cite{EV2,EV}, again in the isotropic space-homogeneous case, study initial data leading to concentration phenomena and blow-up in finite time of the $L^\infty$-norm of the solutions. The paper \cite{BE}  by Briant and Einav removes the isotropy restriction and obtain in polynomially weighted spaces of $L^1\cap L^\infty$ type, existence and uniqueness on a time interval $[0,T_0)$. In \cite{BE} either $T_0=\infty$, or for finite $T_0$ the $L^\infty$-norm of the solution tends to infinity, when time tends to $T_0$. Finally {the space-dependent problem is considered in \cite{R} for a particular setting close to equilibrium, and well-posedness and convergence to equilibrium are proven}.
\\
\hspace*{1.in}\\
The present paper studies a space-dependent, large data problem for the BN equation. The analysis is based on local in time estimates of the mass density.\\
\hspace*{1.in}\\
\setcounter{theorem}{0}
The kernel $B(|v-v_*|,n)$ is assumed measurable with
\begin{equation}\label{hyp1-B}
0\leq B\leq B_0,
\end{equation}
for some $B_0>0$. It is also assumed to depend only on $\lvert v-v_*\rvert $ and $\frac{v-v_*}{\lvert v-v_*\rvert }\cdot n$ denoted by $\cos \theta $, and for some $\gamma >0$, that
\begin{equation}\label{hyp2-B}
B(|v-v_*|, n)=0 \quad \text{for}\hspace*{0.2in} \lvert \cos \theta \rvert <\gamma \hspace*{0.2in}\text{or}\hspace*{0.2in} \lvert 1-\cos \theta \rvert <\gamma .
\end{equation}
These strong cut-off conditions on $B$ are made for mathematical reasons and assumed throughout the paper. For a more general discussion of cut-offs in the collision kernel $B$, see \cite{Lu2}. Notice that contrary to the classical Boltzmann operator where rigorous derivations of $B$ from various potentials have been made, little is known about collision kernels in quantum kinetic theory (cf \cite{V}).\\
Denote by
\begin{equation}\label{f-sharp}
f^{\sharp }(t,x,v)= f(t,x+tv,v)\quad (t,x,v)\in \R _+\times \T ^3 \times \R ^3.
\end{equation}
Strong solutions to the Cauchy problem with initial value $f_0$ associated to the Boltzmann Nordheim equation (\ref{f-bn3D}) are considered in the following sense.
\begin{definition}\label{strong-solution}
$f$ is a strong solution to (\ref{f-bn3D}) on the time interval $I$ if
\begin{eqnarray*}
f\in\mathcal{C}^1(I;L^1(\T ^3\times \R^3)),
\end{eqnarray*}
and
\begin{equation}\label{eq-along-characteristics}
\frac{d}{dt}f^{\sharp }= \big( Q(f)\big) ^{\sharp },\quad \text{on   } I\times \T ^3 \times \R ^3.
\end{equation}
\end{definition}
The main result of this paper is the following.
\begin{theorem}\label{main-theorem}
Assume (\ref{hyp1-B})-(\ref{hyp2-B}). Let $f_0\in L^\infty _+(\T ^3\times \R^3)$ and satisfy
\begin{equation}\label{hyp-f0}
(1+|v|^2)f_0(x,v) \in L^1(\T ^3\times \R ^3), \hspace{.1cm}
\int (1+\lvert v\rvert ^2)\sup_{x\in \T ^3} f_0(x,v)dv=c_0 <\infty .
\end{equation}
There exist a time $T_\infty >0$ and a strong solution $f$ to (\ref{f-bn3D}) on $[0,T_\infty)$ with initial value $f_0$. \\
For $0<T<T_\infty$, it holds
\begin{equation}\label{regularity-f}
f^\sharp \in \mathcal{C}^1([0,T_\infty);L^1(\T ^3\times\R^3))\cap L^\infty ([0,T]\times \T ^3\times \R^3).
\end{equation}
If $T_\infty <+\infty $ then
\begin{equation}\label{explosion}
\limsup _{t\rightarrow T_\infty }\parallel f(t,\cdot ,\cdot )\parallel _{L^\infty (\T ^3\times \R ^3)}= +\infty .
\end{equation}
The solution is unique, depends continuously in $L^1$ on the initial value $f_0$, and conserves mass, momentum, and energy.
\end{theorem}
\underline{\bf Remarks.}\\
A finite $T_\infty$  may not correspond to a condensation. In the isotropic space-homogeneous case considered in \cite{EV2,EV}, additional assumptions on the concentration of the initial value are considered in order to obtain condensation. \\
Theorem \ref{main-theorem} also holds for the classical Boltzmann equation with a similar proof.\\
\\
{To obtain Theorem 1.1 for the boson Boltzmann-Nordheim equation, we start from a fixed
initial value $f_0$ bounded by $2^L$ with $L\in \N$. We shall prove that there are approximations $(f_\alpha )_{\alpha \in ] 0,1] }$ to (\ref{f-bn3D}) and a time $T>0$ independent of $\alpha$, so that $(f_\alpha )$ is bounded by $2^{L+2}$ on $[0,T]$. We then prove that the limit $f$ of $(f_\alpha )$ when $\alpha \rightarrow 0$  solves the bosonic  Boltzmann-Nordheim Cauchy problem (\ref{f-bn3D}).
Iterating the result from T on, it follows that $f$ exists up to the first time $T_\infty $ when (\ref{explosion}) holds.
\\
{The paper is organized as follows.} In the following section, approximations $(f_\alpha )_{\alpha \in ] 0,1] }$ to the Cauchy problem (\ref{f-bn3D}) are constructed. In Section 3 the mass density of $f_\alpha $ is studied with respect to uniform control in $\alpha $. Theorem \ref{main-theorem} is proven in Section 4.
\hspace*{0.1in}\\
%
%
%
%
\section{Approximations.}\label{approx}
\setcounter{equation}{0}
\setcounter{theorem}{0}
{In this section and the following one, the initial datum $f_0$ is assumed to be continuous.}\\
Approximations to the Cauchy problem (\ref{f-bn3D}) are built in the following way. \\
For $\alpha \in ] 0,1] $, let $\chi _\alpha $ be the characteristic function of $[ 0,\frac{1}{\alpha ^2} ] $ and
\begin{align*}\label{df-Ralpha}
R_\alpha (f)(v)= \int_{ I\! \!R^3 \times S^{2}} \chi _\alpha (\lvert v\rvert ^2+\lvert v_*\rvert ^2)B(|v-v_*|,n)
& [ \frac{f^\prime }{1+\alpha f^\prime }\frac{f^\prime _*}{1+\alpha f^\prime _*}\frac{1+f}{1+\alpha f}\frac{1+f_*}{1+\alpha f_*}\nonumber \\
&-\frac{f}{1+\alpha f}\frac{f_*}{1+\alpha f_*}\frac{1+f^\prime }{1+\alpha f^\prime }\frac{1+f^\prime _*}{1+\alpha f^\prime _*}] dv_*dn.
\end{align*}
\begin{lemma}
\hspace*{0.1in}\\
For every $\alpha \in ] 0,1] $, there exists a strong nonnegative space periodic solution
\begin{eqnarray*}
f_\alpha \in \mathcal{C}^1([0,\infty [;L^1(\T ^3\times\R^2))
\end{eqnarray*}
to
\begin{equation}\label{eq-f-alpha}
\partial _tf_{\alpha }+v\cdot \nabla _xf_{\alpha }= R_{\alpha },\quad f_{\alpha }(0,\cdot ,\cdot )= f_0.
\end{equation}
The solution is continuous and unique and conserves mass, momentum and energy.
\end{lemma}
Let $T>0$ be given. We shall first prove by contraction that for $T_1>0$ and small enough, there is a unique solution $f_\alpha $ to (\ref{eq-f-alpha}) on $[0,T_1]$. Let
\begin{eqnarray*}
c_\alpha := \parallel f_0\parallel _\infty +\frac{16\pi ^2B_0T}{3\alpha ^7}\hspace*{0.02in}.
\end{eqnarray*}
Let the map $\mathcal{C}$ be defined on { space periodic functions in}
\begin{eqnarray*}
C \Big( [ 0,T] \times \T ^3 \times \{ v; \lvert v\rvert \leq \frac{1}{\alpha }\} \Big) \cap \{f; f\in [ 0,c_\alpha ] \}
\end{eqnarray*}
by $\mathcal{C}(f)= g$, where $g$ is the unique solution to
\begin{align}\label{eq-g}
&\partial _tg +v\cdot \nabla _xg = \int \chi _\alpha B [ \frac{f^\prime }{1+\alpha f^\prime }\frac{f^\prime _*}{1+\alpha f^\prime _*}\frac{1+f}{1+\alpha f}\frac{1+f_*}{1+\alpha f_*}-\frac{g}{1+\alpha f}\frac{f_*}{1+\alpha f_*}\frac{1+f^\prime }{1+\alpha f^\prime }\frac{1+f^\prime _*}{1+\alpha f^\prime _*}] dv_*dn,\\
&g(0,\cdot ,\cdot )= f_0.\nonumber
\end{align}
It follows from the linearity of the previous partial differential equation that it has a unique {periodic} solution $g$ in $C \big( [ 0,T] \times \T ^3 \times \{ v; \lvert v\rvert \leq \frac{1}{\alpha }\} \big) $. Denote by
\begin{align*}
&R_\alpha ^+(f)(v)= \int \chi _\alpha B\frac{f^\prime }{1+\alpha f^\prime }\frac{f^\prime _*}{1+\alpha f^\prime _*}\frac{1+f}{1+\alpha f}\frac{1+f_*}{1+\alpha f_*}dv_*dn,\\
&\text{and}\\
&\nu _\alpha (f)(v)= \int \chi _\alpha  B\frac{f_*}{1+\alpha f_*}\frac{1+f^\prime }{1+\alpha f^\prime }\frac{1+f^\prime _*}{1+\alpha f^\prime _*}dv_*dn.
\end{align*}
For $f$ nonnegative, $g$ takes its values in $ [ 0,c_\alpha ] $. Indeed,
\begin{eqnarray*}
g^\sharp (t,x,v)\geq f_0(x,v)e^{-\int _0^t( \frac{\nu _\alpha  (f)}{1+\alpha f})^\sharp (s,x,v)ds }\geq 0,
\end{eqnarray*}
and
\[ \begin{aligned}
g^\sharp (t,x,v)&\leq f_0(x,v)+\int_0^t(R_\alpha ^+)^\sharp (s,x,v)ds\leq c_\alpha ,\quad t\in [ 0,T] .
\end{aligned}\]
$\mathcal{C}$ is a contraction in $C\big( [0,T_1] \times \T ^3 \times \{ v;\lvert v\rvert \leq \frac{1}{\alpha }\} \big) \cap \{f; f\in [ 0,c_\alpha  ] \}$, for $T_1>0$ small enough only depending on $\alpha $, since the partial derivatives of the maps
\begin{align*}
(r_i)_{1\leq i\leq 4}\rightarrow \frac{r_3}{1+\alpha r_3}\frac{r_4}{1+\alpha r_4}\frac{1+r_1}{1+\alpha r_1}\frac{1+r_2}{1+\alpha r_2}\text{   and   }(r_i)_{1\leq i\leq 4}\rightarrow \frac{1}{1+\alpha r_1}\frac{r_2}{1+\alpha r_2}\frac{1+r_3}{1+\alpha r_3}\frac{1+r_4}{1+\alpha r_4}
\end{align*}
are bounded on $([ 0,+\infty [ )^4 $ and the domains of integration in $R_\alpha ^+$ and $\nu _\alpha  $ are bounded. Let $f_\alpha $ be its fixed point, i.e. the solution of (\ref{eq-f-alpha}) on $[ 0,T_1] $.\\
 The argument can be repeated and the solution can be continued up to $t= T$.\\
\\
{To obtain Theorem 1.1 for the boson Boltzmann-Nordheim equation, we start from a fixed initial value $f_0$ bounded by $2^L$ with $L\in \N$. We shall prove that there is
a time $T>0$ independent of $\alpha \in ] 0,1] $, so that the solutions $f_\alpha $ to (\ref{eq-f-alpha}) are bounded by $2^{L+2}$ on $[0,T]$. We then prove that the limit $f$ of the solutions $f_\alpha$ when $\alpha\rightarrow 0$  solves the corresponding bosonic  Boltzmann-Nordheim problem.
Iterating the result from T on, it follows that $f$ exists up to the first time $T_\infty$ when
\begin{eqnarray*}
\limsup _{t\rightarrow T_\infty}  \parallel f_\alpha (t,\cdot ,\cdot )\parallel _{L^{\infty }(\T ^3 \times \R ^3)} = \infty .
\end{eqnarray*}
\\
\\
%
%
We observe that
\begin{lemma}\label{T-dependent-on-alpha}
\hspace{.1cm}\\
Given $f_0\leq2^L$ and satisfying (\ref{hyp-f0}), there is for each $\alpha \in ]0,1] $ a time $T_\alpha >0$ so that the solution $f_\alpha$ to (\ref{eq-f-alpha}) is bounded by $2^{L+2}$ on $[0,T_\alpha ]$.
\end{lemma}
\underline{Proof of Lemma \ref{T-dependent-on-alpha}.}\\
Denote $f_\alpha $ by $f$ for simplicity. It holds that
\begin{align}\label{f-sharp}
&\sup_{s\leq t}f^\sharp (s,x,v)
\leq  f_0(x,v)+\int_0^tR_\alpha ^+(f)(s,x+sv,v)ds \nonumber \\
&= f_0(x,v)+\int_0^t\int \chi _\alpha B\frac{f^\sharp }{1+\alpha f^\sharp }(s,x+s(v-v^\prime ),v')\nonumber \\
&\frac{f^\sharp }{1+\alpha f^\sharp }(s,x+s(v-v^\prime _{*}),v'_*)\frac{1+f}{1+\alpha f}(s,x+sv,v)\frac{1+f}{1+\alpha f}(s,x+sv,v_*)dv_*dnds.
\end{align}
Consequently,
\begin{align}\label{bdd-f-sharp}
&\sup_{s\leq t}f^\sharp (s,x,v)\leq  f_0(x,v)+\frac{t}{\alpha ^2}\int B\sup _{(s,x)\in [0,t] \times \T ^3}f^\sharp (s,x,v')\sup _{(s,x)\in [0,t] \times \T ^3}f^\sharp (s,x,v'_*)dv_*dnds.
\end{align}
With the change of variables $(v,v_*,n)\rightarrow (v^\prime ,v_*^\prime, -n)$,
\begin{align}\label{M1-1}
 &\int \sup_{(s,x)\in [0,t]\times \T ^3}f^\sharp (s,x,v)dv
\leq  c_0+\frac{ct}{\alpha ^2}\Big( \int \sup_{(s,x)\in [0,t]\times \T ^3}f^\sharp (s,x,v)dv\Big) ^2,
\end{align}
where
\begin{eqnarray*}
c_0 \text   {   is defined in (\ref{hyp-f0})}\text{   and   }c= 4\pi B_0\hspace*{0.02in}.
\end{eqnarray*}
Denote by
\begin{eqnarray*}
M_1(t)= \int \sup_{(s,x)\in [0,t] \times \T ^3}f^\sharp (s,x,v)dv.
\end{eqnarray*}
It follows from (\ref{M1-1}) that
\begin{equation}
\frac{ct}{\alpha ^2}M_1^2(t)-M_1(t)+c_0\geq 0,\quad t\in [0,\frac{\alpha ^2}{4c_0c}] .\nonumber
\end{equation}
Hence
\begin{equation}\label{M1-2}
M_1(t)\leq \alpha \frac{\alpha -\sqrt{\alpha ^2-4c_0ct}}{2ct}\quad \text{or}\quad M_1(t)\geq \alpha \frac{\alpha +\sqrt{\alpha ^2-4c_0ct}}{2ct},\quad t\in [0,\frac{\alpha ^2}{4c_0c}] \hspace*{0.02in}.
\end{equation}
Moreover,
\begin{equation}\label{equiv}
\alpha \frac{\alpha -\sqrt{\alpha ^2-4c_0ct}}{2ct}\sim c_0\quad \text{and}\quad \alpha \frac{\alpha +\sqrt{\alpha ^2-4c_0ct}}{2ct}\sim \frac{\alpha ^2}{ct},
\end{equation}
when $t$ is a neighborhood of zero. By the continuity of $M_1$ and the behavior of the bounds (\ref{equiv}), it follows from (\ref{M1-2}) that
\begin{equation}
M_1(t)\leq \alpha \frac{\alpha -\sqrt{\alpha ^2-4c_0ct}}{2ct},\quad t\in [0,\frac{\alpha ^2}{4c_0c}] .\nonumber
\end{equation}
And so,
\begin{equation}
M_1(t)\leq 2c_0,\quad t\in [ 0,\frac{\alpha ^2}{4c_0c}] .
\end{equation}
Coming back to (\ref{f-sharp}),
{using the change of variables $v_*\rightarrow v^\prime $ in the gain term of the right-hand side and denoting its Jacobian by $\beta $} leads to
\begin{align}\label{f-alpha-1}
\parallel f_\alpha (t,\cdot ,\cdot )\parallel _{L^\infty (\T ^3 \times \R ^3)}&\leq 2^L+\tilde{c}\int _0^tM_1(s)\parallel f_\alpha (s,\cdot ,\cdot )\parallel _{L^\infty (\T ^3 \times \R ^3)}ds\nonumber \\
&\leq 2^L+2c_0\tilde{c}\int _0^t\parallel f_\alpha (s,\cdot ,\cdot )\parallel _{L^\infty (\T ^3 \times \R ^3)}ds,\quad t\in [0,\frac{\alpha ^2}{4c_0c}] ,
\end{align}
{where $\tilde{c}= \frac{4\pi B_0 \max|\beta| }{\alpha ^2}$.}
And so,
\begin{align}\label{f-alpha-2}
\parallel f_\alpha (t,\cdot ,\cdot )\parallel _{L^\infty (\T ^3 \times \R ^3)}&\leq 2^L\Big( 1+e^{2c_0\tilde{c}t}\Big) \nonumber \\
&\leq 2^{L+2},\quad t\in \Big[ 0,\min \{ \frac{\alpha ^2}{4c_0c}\hspace*{0.02in}, \frac{ln3}{2c_0\tilde{c}} \} \Big] .
\end{align}
The lemma follows.\cqfd
%
%
%
%
\section{Local control of the phase space density.}
\setcounter{equation}{0}
This section is devoted to obtaining a time $T>0$, such that
\begin{eqnarray*}
\sup _{t\in [0,T],\hspace*{0.02in}x\in \T ^3 }f_\alpha ^\sharp (t,x,v)\leq 2^{L+2},
\end{eqnarray*}
uniformly with respect to $\alpha \in ]0,1] $ when $f_0$ is continuous.
\setcounter{theorem}{0}
\[\]
%
%
\begin{lemma}\label{control-mass-densityA}
\hspace*{0.1in}\\
For $T_\alpha $ such that $f_\alpha (t)\leq 2^{L+2}$, $t\in [0,T_\alpha ]$ and $c_0$ defined in (\ref{hyp-f0}), let
\begin{eqnarray*}
\tilde{T}_\alpha = \min \{ T_\alpha , \frac{1}{\pi c_02^{2L+6}}\} .
\end{eqnarray*}
There is a constant $c_1$ independent of $\alpha $, such that the solution $f_\alpha$ of (\ref{eq-f-alpha}) satisfies
\begin{equation}\label{mass-density}
\int (1+\lvert v\rvert ^2)\sup_{(t,x)\in [0,\tilde{T}_\alpha ] \times \T ^3}f_\alpha^\sharp (t,x,v)dv\leq c_1 \hspace*{0.02in}.
\end{equation}
\end{lemma}
\underline{Proof of Lemma \ref{control-mass-densityA}.} \\
It holds that
\begin{align*}
 &\sup_{s\leq t}f^\sharp (s,x,v)
\leq  f_0(x,v)+\int_0^tR_\alpha ^+(f)(s,x+sv,v)ds = f_0(x,v)\nonumber \\
&+\int_0^t\int \chi _\alpha B\frac{f}{1+\alpha f}(s,x+sv,v')\frac{f}{1+\alpha f}(s,x+sv,v'_*)\frac{1+f}{1+\alpha f}(s,x+sv,v)\frac{1+f}{1+\alpha f}(s,x+sv,v_*)dv_*dnds\nonumber\\
&\leq  f_0(x,v)+{2^{2L+6}}\hspace*{0.02in}t\hspace*{0.02in}\int B\sup _{(s,x)\in [0,t] \times \T ^3}f^\sharp (s,x,v')\sup _{(s,x)\in [0,t] \times \T ^3}f^\sharp (s,x,v'_*)dv_*dn.
\end{align*}
With the change of variables $(v,v_*,n)\rightarrow (v^\prime ,v_*^\prime, -n)$ and (\ref{hyp-f0}),
\begin{align*}
 &\int (1+\lvert v\rvert ^2)\sup_{(s,x)\in [0,t]\times \T ^3}f^\sharp (s,x,v)dv
\leq  c_0+c{2^{2L}}t\Big( \int (1+\lvert v\rvert ^2)\sup_{(s,x)\in [0,t]\times \T ^3}f^\sharp (s,x,v)dv\Big) ^2,
\end{align*}
where $c= 2^8\pi B_0$. Denote by
\begin{eqnarray*}
M_2(t)= \int (1+\lvert v\rvert ^2)\sup_{(s,x)\in [0,t] \times \T ^3}f^\sharp (s,x,v)dv.
\end{eqnarray*}
It follows from
\begin{eqnarray*}
c\hspace*{0.02in}2^{2L}t\hspace*{0.02in}M_2^2(t)-M_2(t)+c_0\geq 0,\quad t\in [0,\frac{1}{c_0c\hspace*{0.02in}2^{2L+2}}] ,
\end{eqnarray*}
that
\begin{equation}\label{M2-1}
M_2(t)\leq \frac{1-\sqrt{1-c_0c\hspace*{0.02in}2^{2L+2}t}}{c\hspace*{0.02in}2^{2L+1}t}\quad \text{or}\quad M_2(t)\geq \frac{1+\sqrt{1-c_0c\hspace*{0.02in}2^{2L+2}t}}{c\hspace*{0.02in}2^{2L+1}t}\hspace*{0.02in},\quad t\in [0,\frac{1}{c_0c\hspace*{0.02in}2^{2L+2}}] .
\end{equation}
By the continuity of $M_2$ and the behavior of the bounds
\begin{eqnarray*}
\frac{1-\sqrt{1-c_0c\hspace*{0.02in}2^{2L+2}t}}{c\hspace*{0.02in}2^{2L+1}t}\sim c_0\quad \text{and}\quad \frac{1+\sqrt{1-c_0c\hspace*{0.02in}2^{2L+2}t}}{c\hspace*{0.02in}2^{2L+1}t}\sim \frac{1}{c2^{2L}t}
\end{eqnarray*}
for $t$ in a neighborhood of zero, it follows from (\ref{M2-1}) that
\begin{eqnarray*}
M_2(t)\leq \frac{1-\sqrt{1-c_0c\hspace*{0.02in}2^{2L+2}t}}{c\hspace*{0.02in}2^{2L+1}t},\quad t\in \Big[ 0,\frac{1}{c_0c\hspace*{0.02in}2^{2L+2}}\Big] .
\end{eqnarray*}
And so,
\begin{equation}
M_2(t)\leq 2c_0,\quad t\in [0,\frac{1}{c_0c\hspace*{0.02in}2^{2L+2}}] .
\end{equation}
Bounds on
\begin{eqnarray*}
\int (1+\lvert v\rvert ^2)\sup_{(s,x)\in [\frac{1}{c_0c\hspace*{0.02in}2^{2L+2}}(1+\frac{1}{2}+\cdot \cdot \cdot +\frac{1}{2^n}),\frac{1}{c_0c2^{2L+2}}(1+\frac{1}{2}+\cdot \cdot \cdot +\frac{1}{2^{n+1}})] \times \T ^3}f^\sharp (s,x,v)dv,\quad n\in \N ^*,
\end{eqnarray*}
are analogously obtained by induction. \\
And so, $M_2(t)$ is bounded up to the minimum of $T_\alpha $ and $\frac{1}{c_0c\hspace*{0.02in}2^{2L+2}}\Big( 1+\frac{1}{2}+\frac{1}{2^2}+\cdot \cdot \cdot \Big) = \frac{1}{c_0c\hspace*{0.02in}2^{2L+1}}$\hspace*{0.02in}. \cqfd \\
\begin{lemma}\label{T-independent-on-alphaA}
\hspace{.1cm}\\
Given $f_0\leq2^L$ and satisfying (\ref{hyp-f0}), there are $c_2$ independent on $\alpha $ and $L$, and $T>0$ so that for all $\alpha \in ] 0,1] $, the solution $f_\alpha$ to (\ref{eq-f-alpha}) is bounded by $2^{L+2}$ and
\begin{eqnarray*}
\int (1+\lvert v\rvert ^2)\sup_{(t,x)\in [0,T] \times \T ^3 }f_\alpha^\sharp (t,x,v)dv
\end{eqnarray*}
is bounded by $c_2$ on $[0,T]$.
\end{lemma}
\underline{Proof of Lemma \ref{T-independent-on-alphaA}.}\\
Given $\alpha$, it follows from Lemma \ref{T-dependent-on-alpha} that the maximum time $T^\prime _\alpha $ for which $f_\alpha \leq 2^{L+2}$ on $[0,T^\prime _\alpha ]$ is positive. Moreover,
\begin{align*}
\sup_{s\leq t}f_\alpha ^{\sharp} (s,x,v)
&\leq  f_0(x,v)+\int_0^tR_{\alpha }^{+}(f_\alpha )(s,x+sv,v)ds\\
&\leq f_0(x,v)+2^{3L+8}\int_0^t\int  B\frac{f_\alpha ^\sharp }{1+\alpha f_\alpha ^\sharp}(s,x+s(v-v^\prime ),v')dv_*dn ds.
\end{align*}
With the angular cut-off (2.2), $v_*  \rightarrow v^\prime $ is a change of variables. Using Lemma \ref{control-mass-densityA}, the functions $f_\alpha $ satisfy} for some constant $\bar{c}$,
\begin{align*}
\sup_{(s,x)\in [0,t] \times \T ^3}f_{\alpha }^{\sharp} (s,x,v)&\leq f_0(x,v)+ {2^{L-1}}\bar{c}\hspace*{0.02in}t\int \sup_{(s,x)\in [0,t] \times \T ^3 }f_{\alpha }(s,x,v')dv'\\
&\leq 2^L+{2^{L-1}}\bar{c}\hspace*{0.02in}c_1\hspace*{0.02in}t,\quad  t\in \Big[ 0, \min \{ T^\prime _\alpha ,\frac{1}{\pi c_02^{2L+6}} \} \Big] \, .
\end{align*}
And so,
\begin{align*}
\sup_{s\leq t}f_\alpha ^{\sharp} (s,x,v)
&\leq  3(2^{L-1}),\quad t\in \Big[ 0, \min \{ T^\prime _\alpha ,\frac{1}{\pi c_02^{2L+6}}, \frac{1}{\bar{c}\hspace*{0.02in}c_1} \} \Big] \, .
\end{align*}
For all $\alpha \in ] 0,1] $, it holds that
\begin{eqnarray*}
T^\prime _\alpha \geq \min \{ \frac{1}{\pi c_0\hspace*{0.02in}2^{2L+6}}, \frac{1}{\bar{c}\hspace*{0.02in}c_1} \} ,
\end{eqnarray*}
else $T^\prime _\alpha $ would not be the maximum time such that $f_\alpha (t)\leq 2^{L+2}$ on $[ 0,T^\prime _\alpha ]$ . Consequently,
\begin{align*}
\sup_{s\leq t}f_\alpha ^{\sharp} (s,x,v)
&\leq  2^{L+2},\quad t\in \Big[ 0,  \min \{ \frac{1}{\pi c_0\hspace*{0.02in}2^{2L+6}}, \frac{1}{\bar{c}\hspace*{0.02in}c_1} \} \Big] \, .
\end{align*}
Let
\begin{eqnarray*}
T= \min \{ \frac{1}{\pi c_0\hspace*{0.02in}2^{2L+6}}, \frac{1}{\bar{c}\hspace*{0.02in}c_1} \} .
\end{eqnarray*}
The proof of Lemma \ref{control-mass-densityA} is made again, with $T_\alpha $ replaced by $T$ and leads to the bound $c_2$ of
\begin{eqnarray*}
\int (1+\lvert v\rvert ^2)\sup_{(t,x)\in [0,T] \times \T ^3}f_\alpha ^{\sharp} (t,x,v)dv.
\end{eqnarray*}
\cqfd
%
%
%
%
\section{Proof of Theorem 1.1.}
\setcounter{equation}{0}
\setcounter{theorem}{0}
\hspace*{0.1in}\\
We first prove the existence and uniqueness of a solution to (\ref{f-bn3D}) under the supplementary assumption that $f_0\in C(\T ^3\times \R ^3)$ .\\
Let us first prove that the sequence $(f_\alpha )$ built in Section \ref{approx} is a Cauchy sequence in \\
$C([0,T]; L^1(\T ^3 \times \R ^3))$ {with $T$ of Lemma \ref{T-independent-on-alphaA}. Denote by $F_\alpha $ the function defined by $F_\alpha (x)= \frac{1+x}{1+\alpha x}$\hspace*{0.02in}. For any $(\alpha _1,\alpha _2)\in ]0,1[ ^2$, the function $g=f_{\alpha_1}-f_{\alpha_2}$ satisfies the equation
\[ \begin{aligned}
\partial _tg+v\cdot \nabla _xg&= \int  \chi _{\alpha _{1}}B(f_{\alpha_1}^{\prime} f_{\alpha_1*}^{\prime } -f_{\alpha_2}^{\prime} f_{\alpha_2 *}^{\prime } )F_{\alpha_1}(f_{\alpha_1})F_{\alpha_1}(f_{\alpha_1 *})dv_*dn \\
&-\int  \chi _{\alpha _1}B(f_{\alpha_1}f_{\alpha_1 *} -f_{\alpha _2}f_{\alpha_2 *})F_{\alpha_1}(f_{\alpha_1}^{\prime } )F_{\alpha_1}(f_{\alpha_1 *}^{\prime } )dv_*dn \\
&+\int  \chi _{\alpha _1}Bf_{\alpha_2}^{\prime } f_{\alpha_2 *}^{\prime } \Big( F_{\alpha_1}(f_{\alpha_1 *})\big( F_{\alpha_1}(f_{\alpha_1})-F_{\alpha_1}(f_{\alpha_2})\big) +F_{\alpha_2}(f_{\alpha_2})\big( F_{\alpha_1}(f_{\alpha_1 *})-F_{\alpha_1}(f_{\alpha_2 *})\big) \Big) dv_*dn \\
&+\int  \chi _{\alpha _1}Bf_{\alpha_2}^{\prime } f_{\alpha_2 *}^{\prime} \Big( F_{\alpha_1}(f_{\alpha_1 *})\big( F_{\alpha_1}(f_{\alpha_2})-F_{\alpha_2}(f_{\alpha_2})\big) +F_{\alpha_2}(f_{\alpha_2})\big( F_{\alpha_1}(f_{\alpha_2 *})-F_{\alpha_2}(f_{\alpha_2 *})\big) \Big) dv_*dn \\
& -\int  \chi _{\alpha _1}Bf_{\alpha_2}f_{\alpha_2 *}\Big( F_{\alpha_1}(f_{\alpha_1 *}^{\prime} )\big( F_{\alpha_1}(f_{\alpha_1}^{\prime } )-F_{\alpha_1}(f_{\alpha_2}^{\prime } )\big) +F_{\alpha_2}(f_{\alpha_2}^{\prime} )\big( F_{\alpha_1}(f_{\alpha_1*}^{\prime } )-F_{\alpha_1}(f_{\alpha_2 *}^{\prime } )\big) \Big) dv_*dn \\
&-\int  \chi _{\alpha _1}Bf_{\alpha_2}f_{\alpha_2 *}\Big( F_{\alpha_1}(f_{\alpha_1*}^{\prime} )\big( F_{\alpha_1}(f_{\alpha_2}^{\prime } )-F_{\alpha_2}(f_{\alpha_2}^{\prime} )\big) +F_{\alpha_2}(f_{\alpha_2}^{\prime} )\big( F_{\alpha_1}(f_{\alpha_2 *}^{\prime} )-F_{\alpha_2}(f_{\alpha_2 *}^{\prime} )\big) \Big) dv_*dn\\
&+\int (\chi _{\alpha _1}-\chi _{\alpha _2})\Big(  f_{\alpha_2}^{\prime}f_{\alpha_2 *}^{\prime }F_{\alpha_2}(f_{\alpha_2})F_{\alpha_2}(f_{\alpha_2 *})-f_{\alpha_2}f_{\alpha_2 *}F_{\alpha_2}(f_{\alpha_2}^\prime )F_{\alpha_2}(f_{\alpha_2 *}^\prime )\Big) dv_*dn.\nonumber \\
&\hspace{15cm} (4.8)
\end{aligned}\]
Using Lemma \ref{T-independent-on-alphaA} and taking $\alpha_1<\alpha_2$,
\[ \begin{aligned}
\int   \chi _{\alpha _1}B&\Big(\lvert f_{\alpha_1}f_{\alpha_1*} -f_{\alpha_2}f_{\alpha_2 *}\rvert F_{\alpha_1}(f_{\alpha_1}^{\prime} )F_{\alpha_1}(f_{\alpha_1*}^{\prime} )\Big)^\sharp dxdvdv_*dn \\
&\leq c\hspace*{0.02in}2^{2L}\Big( \int \sup _{x\in \T ^3 }f_{\alpha_1}^{\sharp} (t,x,v)dv
+ \int \sup_ {x\in \T ^3 }f_{\alpha_2}^{\sharp} (t,x,v)dv\Big) \int \lvert (f_{\alpha_1} -f_{\alpha_2} )^{\sharp }(t,x,v)\rvert dxdv\\
&\leq c\hspace*{0.02in}c_22^{2L}\int \lvert g^\sharp(t,x,v)\rvert dxdv.
\end{aligned}\]
We similarly obtain
\begin{eqnarray*}
\int  \chi _{\alpha _1}B \Big( f_{\alpha_2}^{\prime} f_{\alpha_2 *}^{\prime} F_{\alpha_1}(f_{\alpha_1 *})\lvert ( F_{\alpha_1}(f_{\alpha_2})-F_{\alpha_2}(f_{\alpha_2})\lvert )\Big) ^\sharp dxdvdv_*dn \leq c\hspace*{0.02in}c_22^{2L}|\alpha_1-\alpha_2| ,
\end{eqnarray*}
and
\begin{eqnarray*}
\int  \chi _{\alpha _1}B\Big(f_{\alpha_2}f_{\alpha_2 *} F_{\alpha_1}(f_{\alpha_1 *}^{\prime} )\lvert F_{\alpha_1}(f_{\alpha_1}^{^\prime} )-F_{\alpha_1}(f_{\alpha_2}^{\prime} )\rvert \Big)^\sharp dxdvdv_*dn \leq c\hspace*{0.02in}c_22^L\int|g^\sharp(t,x,v)|dxdv.
\end{eqnarray*}
Moreover,
\begin{align*}
&\lvert \int (\chi _{\alpha _1}-\chi _{\alpha _2})\Big(  f_{\alpha_2}^{\prime}f_{\alpha_2 *}^{\prime }F_{\alpha_2}(f_{\alpha_2})F_{\alpha_2}(f_{\alpha_2 *})-f_{\alpha_2}f_{\alpha_2 *}F_{\alpha_2}(f_{\alpha_2}^\prime )F_{\alpha_2}(f_{\alpha_2 *}^\prime )\Big) dxdvdv_*dn\rvert \\
&\leq c\hspace*{0.02in}2^{2L}\int _{\mid v\mid >\frac{1}{\sqrt{2}\alpha _1}\hspace*{0.02in}\text{or}\hspace*{0.02in}\mid v_*\mid >\frac{1}{\sqrt{2}\alpha _1}}f_{\alpha_2}(t,x,v)f_{\alpha_2 }(t,x,v_*)dxdvdv_*\\
&\leq c\hspace*{0.02in}c_22^{2L}\int _{\mid v\mid >\frac{1}{\sqrt{2}\alpha _1}}f_{\alpha_2}(t,x,v)dxdv\\
&\leq c\hspace*{0.02in}c_22^{2L}\alpha _1^2\int \lvert v\rvert ^2f_{\alpha _2}(t,x,v)dxdv.
\end{align*}
The remaining terms are estimated in the same way. It follows
\begin{eqnarray*}
\frac{d}{dt}\int|g^\sharp(t,x,v)|dxdv\leq c\hspace*{0.02in}c_22^{2L}\Big(\int|g^\sharp (t,x,v)|dxdv+|\alpha_1-\alpha_2| +\alpha _1^2\Big).
\end{eqnarray*}
 Hence
 \begin{eqnarray*}
 \lim _{(\alpha _1,\alpha _2)\rightarrow (0,0)}\sup _{t\in [0,T]}\int |g^\sharp(t,x,v)|dxdv= 0.
 \end{eqnarray*}
 And so $(f_\alpha )$ is a Cauchy sequence in $C([0,T]; L^1(\T ^3 \times \R ^3))$. Denote by $f$ its limit. Analogously,
 \begin{eqnarray*}
 \lim _{\alpha \rightarrow 0}\int \lvert Q(f)-Q(f_\alpha )\rvert (t,x,v)dtdxdv= 0.
 \end{eqnarray*}
 Hence $f$ is a strong solution to (\ref{f-bn3D}) on $[0,T]$ with initial value $f_0$. \\
 If there were two solutions, their difference denoted by $G$ would with similar arguments satisfy
 \begin{eqnarray*}
\frac{d}{dt}\int |G^\sharp(t,x,v)|dxdv\leq c\hspace*{0.02in}c_22^{2L}\int |G^\sharp(t.x.v)|dxdv,
 \end{eqnarray*}
 hence be identically equal to its initial value zero. And so there exists a unique solution to (\ref{f-bn3D}).
 \hspace*{0.1in}\\
 \hspace*{0.1in}\\
 Let us prove the existence and uniqueness of a solution to (\ref{f-bn3D}) for any {$f_0\in L^\infty _+(\T ^3\times \R ^3)$} satisfying (\ref{hyp-f0}).\\
 If $f_1$ (resp. $f_2$) is a solution to (\ref{f-bn3D}) with the continuous initial value $f_{10}$ (resp. $f_{20}$), then similar arguments lead to
\begin{eqnarray*}
\frac{d}{dt}\int |(f_1-f_2)^\sharp(t,x,v)|dxdv\leq c\hspace*{0.02in}c_22^{2L}\int |(f_1-f_2)^\sharp(t,x,v)|dxdv,
 \end{eqnarray*}
 so that
 \begin{eqnarray*}
\parallel (f_1-f_2)(t,\cdot ,\cdot )\parallel _{L^1(\T ^3 \times \R ^3)}\leq e^{c\hspace*{0.02in}c_22^{2L}T}\parallel f_{10}-f_{20}\parallel _{L^1(\T ^3 \times \R ^3)},\quad t\in [0,T] .
 \end{eqnarray*}
 Consider $f_0$ as the limit in $L^1(\T ^3 \times \R ^3)$ of a sequence $(f_{0,n})_{n\in \N }$ of continuous functions satisfying (\ref{hyp-f0}). Let $(f_n)_{n\in \N }$ be the solutions to (\ref{f-bn3D}) associated to the initial data $(f_{0,n})_{n\in \N }$. It can similarly be proven that $(f_n)_{n\in \N }$ is a Cauchy sequence in $C([0,T]; L^1(\T ^3\times \R ^3))$ and that its limit $f$ is the unique solution in
$C([0,T]; L^1(\T ^3\times \R ^3))$ to (\ref{f-bn3D}).\\
\hspace*{1.in}\\
 Finally, if $f_1$ (resp. $f_2$) is the solution to (\ref{f-bn3D}) with initial value $f_{10}$ (resp. $f_{20}$), then similar arguments lead to
\begin{eqnarray*}
\frac{d}{dt}\int |(f_1-f_2)^\sharp(t,x,v)|dxdv\leq c\hspace*{0.02in}c_22^{2L}\int |(f_1-f_2)^\sharp(t,x,v)|dxdv,
 \end{eqnarray*}
 so that
 \begin{eqnarray*}
\parallel (f_1-f_2)(t,\cdot ,\cdot )\parallel _{L^1(\T ^3 \times \R ^3)}\leq e^{c\hspace*{0.02in}c_22^{2L}T}\parallel f_{10}-f_{20}\parallel _{L^1(\T ^3 \times \R ^3)},\quad t\in [0,T] ,
 \end{eqnarray*}
i.e. stability holds. \\
\hspace*{1.in}\\
If
\begin{eqnarray*}
\sup_{(x,v)\in \T ^3 \times \R ^3}f(T,x,v)<2^{L+2},
\end{eqnarray*}
then the procedure can be repeated, i.e. the same proof can be carried out from the initial value $f(T)$. It leads to a maximal interval denoted by $[0,\tilde {T}_1]$ on which $f(t,\cdot ,\cdot )\leq 2^{L+2}$. By induction there exists an increasing  sequence of times $(\tilde{T}_n)$  such that $f(t,\cdot ,\cdot )\leq 2^{L+2n}$ on $[ 0,\tilde{T}_n]$. Let
\begin{eqnarray*}
\tilde{T}_\infty = \lim _{n\rightarrow +\infty }\tilde{T}_n.
\end{eqnarray*}
Either $\tilde{T}_\infty = +\infty $ and the solution $f$ is global in time, or $\tilde{T}_\infty $ is finite and then the {limes superior of the solution is infinity} in the $L^\infty$-norm at $\tilde{T}_\infty$. \cqfd
 %
 %
%
%
%
\begin{lemma}\label{conservations}
The solution $f$ to (\ref{f-bn3D}) with initial value $f_0$ conserves mass, momentum and energy.
\end{lemma}
\underline{Proof of Lemma  \ref{conservations}.}\\
The conservation of mass and first momentum of $f$ will follow from the boundedness of the total energy. The energy is non-increasing since the approximations $f_\alpha $ conserve energy and
\begin{eqnarray*}
\lim _{\alpha \rightarrow 0}\int _{\T ^3}\int _{\lvert v\rvert <V}\lvert (f-f_\alpha )(t,x,v)\rvert \lvert v\rvert ^2dxdv= 0,\quad \text{for all }t\in [0,T_\infty[ \text{   and positive  }V.
\end{eqnarray*}
 Energy conservation will be satisfied if the energy is non-decreasing. Taking $\psi_\epsilon=\frac{|v^2|}{1+\epsilon|v|^2}$  as approximation for $|v|^2$, it is enough to bound
\begin{eqnarray*}
\int R_0(f)(t,x,v)\psi_\epsilon (v)dxdv = \int B\psi_{\epsilon}\Big( f^\prime f^\prime _{*}(1+f)(1+f_{*})
- ff_{*}(1+f^\prime )(1+f^\prime _{*})\Big) dxdvdv_*dn
\end{eqnarray*}
from below by zero in the limit $\epsilon \rightarrow 0$. Similarly to{ \cite{Lu2}},
\[ \begin{aligned}
\int R_0(f)\psi_\epsilon dxdv
&=\frac{1}{2}\int B ff_{*}(1+f^\prime )(1+f^\prime _{*})\Big( \psi_\epsilon(v')+\psi_\epsilon(v'_*)-\psi_\epsilon(v)-\psi_\epsilon(v_*)
\Big)dxdvdv_*dn \\
&\geq -\int Bff_{*}(1+f^\prime )(1+f^\prime _{*})\frac{\epsilon |v|^2|v_*|^2}{(1+\epsilon|v|^2)(1+\epsilon|v_*|^2)}dxdvdv_*dn\\
&\geq -c\hspace*{0.02in} \epsilon \hspace*{0.02in}2^{2L}\int \lvert v\rvert ^2\sup _{(t,x)\in [0,T] \times \T ^3}f^\sharp (t,x,v)dv \int \lvert v_*\rvert ^2f(t,x,v_*)dxdv_*\\
&\geq -c \hspace*{0.02in}c_2\hspace*{0.02in} \epsilon \hspace*{0.02in}2^{2L}.
\end{aligned}\]
This implies that the energy is non-decreasing, and bounded from below by its initial value. That completes the proof of the lemma.     \cqfd
\\
\\
\\

\[\]

\end{document}